\documentclass[USenglish,oneside,twocolumn,notitlepage]{article}
\usepackage{etex}
\reserveinserts{28}

\usepackage{colortbl}
\usepackage{multirow}
\usepackage{balance}
\usepackage{comment}
\usepackage[nolist]{acronym}
\usepackage{xspace}
\usepackage{float}
\usepackage{lipsum}
\usepackage{dblfloatfix}
\usepackage[numbers]{natbib}
\usepackage{amsmath}

\definecolor{gray}{rgb}{.85,.85,.85}
\definecolor{yellow}{rgb}{.98,.98,.82}
\definecolor{Gray}{gray}{0.9}
\definecolor{LightCyan}{rgb}{0.88,1,1}

\newcommand*{\ti}{\hspace*{0.2cm}}%

\usepackage[noend]{algpseudocode}
\usepackage{algorithm}
\usepackage{soul}
\usepackage{color}
\usepackage[table,x11names]{xcolor}
\usepackage{pgfplots}
\usepackage{tikz}
\usepackage[tight]{subfigure}
\usepackage{verbatim}

\usepackage{adjustbox}
\usepackage{listings}

\usepackage{tikz}
\usetikzlibrary{calc,trees,positioning,arrows,chains,shapes.geometric,%
    decorations.pathreplacing,decorations.pathmorphing,shapes,%
    matrix,shapes.symbols}

\tikzstyle{state} = [rectangle, rounded corners, draw=black, very thick, text width=10em, text centered]
\tikzstyle{line} = [draw, thick, <-]

\graphicspath{{./figs/}}

\usepackage[utf8]{inputenc}

\usepackage[justification=centering]{caption}
\usepackage[hyphens]{url}
\newcounter{bob}

\newcommand{\bi}{\begin{itemize}}
\newcommand{\ei}{\end{itemize}}
\newcommand{\lt}{layer-2\@\xspace}

\usepackage{authblk}

\title{A Study of MAC Address Randomization in Mobile Devices and When it Fails}

\author{Jeremy Martin \thanks{jmartin@usna.edu}} 
\author{Travis Mayberry \thanks{mayberrry@usna.edu}}         
\author{Collin Donahue}          
\author{Lucas Foppe}         
\author{Lamont Brown}          
\author{Chadwick Riggins}          
\author{Erik C. Rye \thanks{rye@usna.edu}}
\author{Dane Brown \thanks{dabrown@usna.edu}}
\affil{US Naval Academy} 

\date{\vspace{-5ex}}
\begin{document}   

\maketitle

\begin{abstract}
{\ac{MAC} address randomization is a privacy technique whereby mobile devices
  rotate through random hardware addresses in order to prevent observers from
  singling out their traffic or physical location from other nearby devices.
  Adoption of this technology, however, has been sporadic and varied across device
  manufacturers.  In this paper, we present the first wide-scale study of
  \ac{MAC} address randomization in the wild, including a detailed breakdown of
  different randomization techniques by operating system, manufacturer, and model of
  device.\\ We then identify multiple flaws in these implementations which can be exploited to defeat randomization as
  performed by existing devices.  First, we show that devices commonly make
  improper use of randomization by sending wireless frames with the true, global
  address when they should be using a randomized address.  We move on to extend
  the passive identification techniques of Vanhoef et al. to effectively defeat
  randomization in $\sim$96\% of Android phones.  Finally, we show a method that can be used to track 100\% of devices using randomization,
  regardless of manufacturer, by exploiting a previously unknown flaw in the way existing wireless
  chipsets handle low-level control frames.}
\end{abstract}




\section{Introduction}
\label{sec:intro}

Smartphones are one of the most impactful technologies of this century.  The
ability to access the Internet anytime and anywhere has fundamentally changed
both work and personal life across the globe \cite{sarwar2013impact}.  It is
gradually becoming clear, however, that in exchange for this level of access to
the Internet people may be giving up a substantial amount of privacy.  In
particular, it has recently been made public that state sponsored intelligence
agencies, in countries such as Russia and China \cite{intelligence_2017,
darakerr_2013, foundation_2011}, as well as private sector companies
\cite{mims_2012}, are actively attempting to track cellphone users.

Smartphones conventionally have two major modes of communication, both of which
can potentially be used to track users.  The first and most obvious is the
cellular radio itself \cite{bard2016unpacking,owsley2015spies}.
However, an often overlooked second avenue for tracking cellphones (and their
corresponding users) is the 802.11 (WiFi) radio that most smart phones also use.

Every 802.11 radio on a mobile device possesses a 48-bit link-layer \ac{MAC}
address that is a globally unique identifier for that specific WiFi device.  The
\ac{MAC} address is a crucial part of WiFi communication, being included in
every link-layer frame that is sent to or from the device.  This unfortunately
poses a glaring privacy problem because any third party eavesdropping on nearby
WiFi traffic can uniquely identify nearby cellphones, and their traffic, through
their \ac{MAC} addresses \cite{Cunche2014}.

There is one particular type of WiFi packet, called a \emph{probe request
frame}, that is an especially vulnerable part of WiFi traffic with respect to
surveillance. Since probe requests continuously broadcast at a semi-constant
rate they make tracking trivial.  Mobile devices are effectively playing an
endless game of digital ``Marco Polo,'' but in addition to ``Marco'' they are
also broadcasting out their IDs (in the form of a \ac{MAC} address) to anyone
that cares to listen. To address this problem, some modern mobile devices make
use of temporary, randomized \ac{MAC} addresses that are distinct from their
true global address.  When probe requests are sent out, they use a randomized
\emph{pseudonym} \ac{MAC} address that is changed periodically.  A listener
should be unable to continuously track the phone because the \ac{MAC} changes in
a way that hopefully cannot be linked to the previous address.

\textbf{In this work we evaluate the effectiveness of various deployed
\ac{MAC} address randomization schemes}.  We first investigate how
exactly different mobile \acp{OS} actually implement randomization
techniques, specifically looking at how the addresses are generated
and under what conditions the devices actually use the randomized
address instead of the global one.  Using real-world datasets we
provide the first evaluation of adoption rates for randomization across
a diverse manufacturer and model corpus.

After establishing the current state of randomization for widely used
phone models and OS versions, we move on to show several weaknesses in
these schemes that allow us to track phones within and across multiple
collections of WiFi traffic.  Our work builds on the fingerprinting
techniques of \citet{vanhoef2} in addition to new approaches for
deanonymizing phones based on weaknesses we discovered while analyzing
wireless traffic from many randomizing phones.

This paper makes the following novel contributions:

\vspace{2mm}

\bi
\item We decompose a large 802.11 corpus, providing the first
  granular breakdown of real-world \ac{MAC} address randomization.
  Specifically, we develop novel techniques to identify and isolate randomization
  and randomization schemes from large collections of wireless traffic.
\item We present the first manufacturer and device breakdown for \ac{MAC} randomization, describing the particular technique each uses.  Our results indicate that adoption rates are surprising low, specifically for Android devices.
\item We review previous techniques for determining global \ac{MAC} addresses and find them to be insufficient. We provide additional context and improvements to existing passive and active techniques, substantially increasing their effectiveness.
\item We identify significant flaws in the majority of
  randomization implementations on Android devices.  These flaws allow for
  trivial retrieval of the global \ac{MAC} address.
\item Discovery and implementation of a control frame attack which exposes the global \ac{MAC}
  address (and thus allows tracking/surveillance) for all known devices, regardless of \ac{OS}, manufacturer, device type, or
  randomization scheme.  Furthermore, Android devices can be
  susceptible to this attack even when the user disables WiFi and/or
  enables Airplane Mode.
    \ei

\section{Background}
\label{sec:background}

\subsection{MAC Addresses}

Every network interface on an 802.11 capable device has a 48-bit \ac{MAC}
address \lt hardware identifier. \ac{MAC} addresses are designed to be
persistent and globally unique.  In order to guarantee the uniqueness of
\ac{MAC} addresses across devices the \ac{IEEE} assigns blocks of addresses to
organizations in exchange for a fee.  A \ac{MA-L}, commonly known as an
\ac{OUI}, may be purchased and registered with the \ac{IEEE}
\cite{ieeeOUIlisting}, which gives the organization control of and
responsibility for all addresses with a particular three-byte prefix.  The
manufacturer is then free to assign the remaining low-order three bytes
($2^{24}$ distinct addresses) any value they wish when initializing devices,
subject to the condition that they do not use the same \ac{MAC} address twice.


An implication of the \ac{IEEE} registration system is that it is trivial to
look up the manufacturer of a device given its \ac{MAC} address.  Using, again,
the example of a wireless eavesdropper, this means that anyone listening to
802.11 traffic can determine the manufacturer of nearby devices.  To combat this,
the \ac{IEEE} also provides the ability to purchase a ``private'' \ac{OUI} which
does not include the company's name in the register.  However, this additional
privacy feature is not currently used by any major manufacturers that we are
aware of.

\begin{figure}[h]
\centering
\begin{tikzpicture} [node distance=.5cm, auto,every node/.style={align=center}]
  \node[rectangle,draw=black] (M1) {\large \texttt{01}};
\node [right of=M1](C1) {\large :};
  \node[rectangle,draw=black, right of = C1] (M2) {\large \texttt{23}};
\node [right of=M2](C2) {\large :};
  \node[rectangle,draw=black, right of = C2] (M3) {\large \texttt{45}};
\node [right of=M3](C3) {\large :};
  \node[rectangle,draw=black, right of = C3] (M4) {\large \texttt{67}};
\node [right of=M4](C4) {\large :};
  \node[rectangle,draw=black, right of = C4] (M5) {\large \texttt{89}};
\node [right of=M5](C5) {\large :};
\node[rectangle,draw=black, right of = C5, shift={(.5mm, 0mm)}] (M6) {\large
  \texttt{AB}};
\draw[decorate, decoration={brace, amplitude=10pt}] (M1.north west) -- (M3.north east) node [above, align=center, midway, yshift=4mm] {\large OUI};
\draw[decorate, decoration={brace, amplitude=10pt}] (M4.north west) -- (M6.north east) node [above, align=center, midway, yshift=4mm] {\large NIC};
\node[rectangle,draw=black, below of = M1, yshift=-5mm, xshift=0mm] (B3)
  {\texttt{000000\textcolor{red}{01}}};
\draw (M1.south west) -- (B3.north west);
\draw (M1.south east) -- (B3.north east);
\node[right=of B3, yshift=-1mm, xshift=2mm] (mc) {Unicast/Multicast Bit};
\node[below=of B3, yshift=-1mm, xshift=3mm] (loc) {Universal/Local Bit};
\draw[<-] ($(B3.east)+(-1.4mm, -1mm)$) -- (mc.west);
\draw[->] ($(loc.north)+(1.4mm,0mm)$) -- ($(B3.east)+(-3.8mm, -1.5mm)$);
\end{tikzpicture}

  \caption{48-bit MAC Address Structure}
  \label{fig:macaddr}
\end{figure}
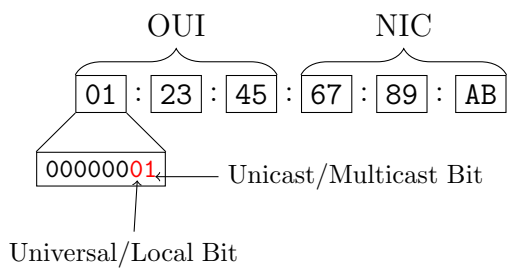


In addition to the public, globally unique, and manufacturer assigned \ac{MAC}
address, modern devices frequently use \emph{locally assigned} addresses
\cite{rfc7042} which are distinguished by a Universal/Local bit in the most
significant byte.  Locally assigned addresses are not guaranteed to be unique,
and generally are not used in a persistent manner.  Locally assigned addresses
are used in a variety of contexts, including multi-\ac{SSID} configured \acp{AP},
mobile device-tethered hotspots, and \ac{P2P} services.  A visual depiction of
the \ac{MAC} address byte structure is illustrated in Figure~\ref{fig:macaddr}.

Most importantly for this paper, locally assigned addresses may also
be used to create randomized \ac{MAC} addresses as an additional measure of
privacy. Similar to an \ac{OUI}, a three-byte \ac{CID} prefix can be
purchased from the IEEE, with the agreement that
assignment from this address space will not be used for globally unique
applications.  As such, a \ac{CID} always has the local bit set, and
is predisposed for use within \ac{MAC} address randomization
schemas. One such example, the Google owned \texttt{DA:A1:19} \ac{CID},
is prominent within our dataset.

With the advent of randomized, locally assigned \ac{MAC} addresses
that change over time, tracking a wireless device is no longer
trivial.  For this reason, we frequently observe 802.11 probe requests
using locally assigned addresses when the device is in a disassociated
state (not associated with an \ac{AP}).  When a mobile device attempts
to connect to an \ac{AP}, however, it reverts to using its globally unique \ac{MAC}
address.  As such, tracking smartphones becomes trivial while
they are operating in an associated state.  

Since mobile devices are usually only associated while the user is relatively
stationary (otherwise they would be out of range of the AP), tracking them in
this state is less of a privacy vulnerability than having the ability to track
devices in an unassociated state, which usually occurs when the user is moving
from one location to another.  Additionally, there are several good reasons to
use a global address in an associated state, such as to support \ac{MAC} address
filtering on the network.  Therefore we concentrate, in this paper, on
evaluating randomization methods and tracking of unassociated devices.

\subsection{Mobile \ac{OS} \ac{MAC} Randomization}

A particularly sensitive privacy issue arises from the manner in which
wireless devices identify access points within close proximity.
Traditionally, devices perform \emph{active scanning} where they broadcast
probe request frames asking nearby \acp{AP} to identify themselves
and respond with 802.11 parameter information required for connection
setup.  These probe request frames require a source \ac{MAC} address,
but if an 802.11 device uses its globally unique \ac{MAC} address then
it is effectively broadcasting its identity at all times to any
wireless receiver that is nearby.  Wireless device users can then
easily be tracked across temporal and spatial boundaries as their
devices are transmitting with their unique identity.

To combat this privacy concern, both Android and Apple iOS operating
systems allow for devices in a disassociated state to use random,
locally assigned \ac{MAC} addresses when performing active scans.  Since
the \ac{MAC} address is now random, users gain a measure of anonymity
up until they associate with an \ac{AP}. 

The particular software hooks used for randomization vary between operating
systems.  See Appendix \ref{wpasup} for a discussion of the OS mechanisms and
configuration files that support \ac{MAC} randomization.

\section{Related Work}
\label{sec:related}

\citet{vanhoefRandomMAC} present several techniques for tracking devices
regardless of privacy countermeasures such as \ac{MAC} address randomization.
These attacks rely on devices' support for \ac{WPS}, a protocol that allows
unauthenticated devices to negotiate a secure connection with access points.
Unfortunately, in order to facilitate this process, extra \ac{WPS} fields are
added in a device's probe requests that contain useful information for device
tracking.  Among these is the manufacturer and model of the device, but also a
unique identifier called the \ac{UUID-E} which is used to establish \ac{WPS}
connections.  The flaw that \citet{vanhoefRandomMAC} discovered is that the
\ac{UUID-E} is derived from a device's global \ac{MAC} address, and by using
pre-computed hash tables an attacker can simply lookup the \ac{UUID-E} from the
table and retrieve the global \ac{MAC} address
\cite{vanhoefRandomMAC,ourpaper2}.  We refer to this technique as
\emph{\ac{UUID-E} reversal}.  Since the \ac{UUID-E} does not change, the
implication is that even if the \ac{MAC} address is randomized, an attacker can
still recover the original, global address by performing this reversal technique
on the \ac{UUID-E}.

While the revelation of the flaw was significant, several holes in the
analysis were observed due to the dataset on which the work was evaluated.  The
attack was applied against an anonymized dataset from 2013
\cite{sapienza-probe-requests-20130910}.  This dataset did not include
randomized \ac{MAC} address implementations as they did not exist in 2013.
Additionally, due to the fact that the data was anonymized, and ground truth was
not available, a validation of the reversal technique was not provided.  The
authors state that the address could not be confirmed to be the WiFi \ac{MAC}
address, rather it may represent the Bluetooth \ac{MAC} address of the device.
Because of this, the reader is left with little understanding on the scope of
practical use of these attacks.  Namely, is the attack truly viable against
devices performing randomization?

The first contribution of this paper is a better evaluation of the attacks
presented by \citet{vanhoef2}. Using more recent real-world data, we verify that
this technique is plausible for defeating randomization for a small set of
devices.  However, we also show that an improvement on their technique can
achieve a higher success rate, up 99.9\% effectiveness against vulnerable
devices.  We are also able to confirm that the retrieved \ac{MAC} address is in
fact the 802.11 WiFi identifier and not the Bluetooth address using additional
techniques.  More importantly, we provide a real-world assessment for the scope
of the attack, revealing that only a small portion of Android devices are
actually vulnerable.

\citet{vanhoefRandomMAC} and \citet{vanhoef2} present an additional technique:
fingerprinting of the probe request 802.11 \acp{IE}.  \acp{IE} are optional,
variable length fields which appear in WiFi management frames and are generally
used to implement extensions and special features on top of the standard WiFi
protocol.  Importantly, there are enough of these extensions and manufacturer
specific functions that the various combinations which are supported on a
particular device may be unique to that device, causing the \acp{IE} to form a
fingerprint which can be used to identify traffic coming from that device.  

However, we find one significant flaw in the evaluation of these fingerprints:
locally assigned \ac{MAC} addresses were ignored by the authors.  Nearly all randomization
schemes utilize locally assigned \ac{MAC} addresses to perform randomization.
As such, previous research failed to identify problems observed when
tracking randomized \ac{MAC} addresses.  A simple example of this is the
signature of a device's probe request, which we observed changing during
randomization and even when not randomizing.  Only by observing these behaviors
can we truly implement effective derandomization techniques and present honest
reflections on the limitations of the attack methods. 

Also presented in \cite{vanhoefRandomMAC} is a revival of the Karma attack using
a top-n popular \ac{SSID} honeypot approach.  As noted above, \ac{MAC}
randomization stops once a device becomes associated with an \ac{AP}.  Karma
attacks are active attacks where a rogue \ac{AP} is configured with an identical
name (\ac{SSID}) to one that the device is set up to automatically connect to.
In effect, this forces the devices into an authenticated state where it reveals
its global \ac{MAC} address and bypass randomization.  We validate this attack
by finding that the increased prevalence of seamless WiFi-offloading from
cellular networks means that many devices in the wild are vulnerable.

\section{Methodology}
\label{sec:methodology}

Our initial goal is to identify which mobile devices are using randomization, in
order to narrow down further investigation into their exact methods for doing
so.  Since this is not a capability that is advertised on a spec sheet, we
resort to broad capture and analysis of WiFi traffic in order to determine which
device models are doing randomization.

Over the course of approximately two years, we captured unencrypted
802.11 device traffic using inexpensive commodity hardware and
open-source software. We primarily use an LG Nexus 5 Android phone
running Kismet \emph{PcapCapture} paired with an AWUS036H 802.11b/g
Alfa card.  We hop between the 2.4GHz channels 1, 6, and 11 to
maximize coverage. We additionally employ several Raspberry Pi devices
running Kismet with individual wireless cards each dedicated to
channels 1, 6, and 11.  Our corpus spans January 2015 to December 2016
and encompasses approximately 9,000 individual packet captures.  The
collection contains over 600 \acp{GB} of 802.11 traffic, consisting of
over 2.8 million unique devices.

It is important to note that, since devices only randomize when they are
unassociated, the only traffic we are interested in is 802.11 management frames
and unencrypted \ac{mDNS} packets.  Therefore we did not capture actual
intentional user traffic from the device, i.e. web browsing, email, etc., but only
automatic, non-personal traffic sent by the device.

\subsection{Ethical Considerations}
\label{sec:irb}
Our collection methodology is entirely passive.  At no time did we attempt to
decrypt any data, or perform active actions to stimulate or alter normal network
behavior while outside of our lab environment.  Our intent is to show the ease
with which one can build this capability with low-cost, off-the-shelf equipment.
However, given the nature of our data collection, we consulted with our
\ac{IRB}.

\begin{table*}[t]
	\scriptsize
    
\adjustbox{valign=t}{\begin{minipage}[c]{0.3\textwidth}
 \caption{Corpus Statistics}
  \vspace{-2mm}
  \label{table:probes}
 \begin{tabular}{ l | l }\hline  
  \textbf{Category}  & \textbf{\# \acp{MAC} } \\ \hline \hline
      Corpus              & 2,604,901 \\ \hline
    \ti Globally Unique   & \ti 1,204,148 \\ \hline
    \ti Locally Assigned  & \ti 1,400,753 \\ \hline
  \end{tabular}
\end{minipage}}
\adjustbox{valign=t}{\begin{minipage}[c]{0.3\textwidth}
 \caption{Locally Assigned Bins}
  \vspace{-2mm}
  \label{table:local}
 \begin{tabular}{ l | l }\hline  
  \textbf{Category}  & \textbf{\# \acp{MAC}} \\ \hline \hline
     Locally Assigned  & 1,400,753 \\ \hline
        \ti Service    & \ti 3,147     \\ \hline
        \ti Randomized & \ti 1,388,566  \\ \hline
  	   	\ti Unknown    & \ti 9,040       \\  \hline 
	\end{tabular}
\end{minipage}}
\adjustbox{valign=t}{\begin{minipage}[c]{0.3\textwidth}
 \caption{Randomization Bins}
  \vspace{-2mm}
  \label{table:random}
 \begin{tabular}{ l | l }\hline  
  \textbf{Category}  & \textbf{\# \acp{MAC} } \\ \hline \hline
        Randomized       &  1,388,566  \\ \hline
        \ti Android: \texttt{DA:A1:19} (WPS)  &  \ti 8,761\\ \hline
        \ti Android: \texttt{DA:A1:19}  &  \ti 43,924\\ \hline
        \ti Android: \texttt{92:68:C3} (WPS)  &  \ti 8,961\\ \hline
        \ti iOS                &  \ti 1,326,951 \\ \hline
        \ti Windows 10 / Linux       &  \ti 59        \\ \hline

  \end{tabular}
\end{minipage}}
\end{table*}

The primary concerns of the \ac{IRB} centered on: i) the information
collected; and ii) whether the experiment collects data ``about whom''
or ``about what.''  Because we limit our analysis to 802.11
management frames and unencrypted \ac{mDNS} packets, we do not observe
\ac{PII}.  Although we observe IP addresses, our experiment does not
use these layer-3 addresses.  Even with an IP address, we have no
reasonable way to map the address to an individual.  Further, humans
are incidental to our experimentation as our interest is in the
randomization of wireless device layer-2 \ac{MAC} addresses, or ``what.''
Again, we have no way to map \ac{MAC} addresses to individuals.

Finally, in consideration of beneficence and respect for persons, our
work presents no expectation of harm, while the concomitant
opportunity for network measurement and security provides a societal
benefit.  Our experiment was therefore determined to not be human
subject research.

\subsection{Identifying Randomization}
\label{sec:methRand}

We know devices implement \ac{MAC} randomization in different ways.  In order to
quantify the vulnerabilities of employed randomization policies, we first
attempt to categorize devices into different \emph{bins}, with identical
behavior, so that we can investigate characteristics of these individual
techniques and seek to identify flaws in their implementation.  For instance, as
we will see, all iOS devices fall into the same bin, in that they handle
randomization in a similar way.  Android devices, on the other hand, differ
significantly from iOS, and also vary greatly from manufacturer to manufacturer.

Our first step is to identify whether a device is performing randomization.
This starts with extracting all source \ac{MAC} addresses derived from probe
request frames in our corpus.  If the local bit of the \ac{MAC} address is set,
we store the address as a locally assigned \ac{MAC} address in our database.
Since randomized addresses cannot be unique, we assume at this point that any
device using randomization will set the local bit in its \ac{MAC} address and
therefore all randomization candidates will be in this data set.  For each
address we then parse the advertised \ac{WPS} \texttt{manufacturer},
\texttt{model\_name}, \texttt{model\_number}, and \texttt{uuid\_e} values.
Additionally, we build signatures derived from a mapping of the advertised
802.11 \ac{IE} vendor fields using techniques from related work in device-model
classification \cite{vanhoefRandomMAC, googlepaper}.  Each \ac{MAC}
address, associated \ac{WPS} values (when applicable), and the device \ac{IE}
signature are stored in our database.

Our device signatures are created using custom built Wireshark dissectors to
parse the 802.11 vendor \ac{IE} fields and values.  Our modifications to
standard wireshark files (\emph{packet-ieee80211.c} and
\emph{packet-ieee80211.h}) allow us to efficiently create the individual device
signatures as we process the packet captures, eliminating any need for
post-processing scripts.  Furthermore, this allows us to use a signature as a
display filter while capturing.  We will later use the device signatures for
both passive and active derandomization techniques.

Our corpus contained a total of $\sim$66 million individual probe requests.  We
have a dataset of 2.6 million unique source \ac{MAC} addresses after removing
duplicates.  In Table~\ref{table:probes} we observe that 1.4 million
($\sim$53\%) of the 2.6 million distinct \ac{MAC} addresses had locally assigned
\ac{MAC} addresses.  Recall that locally assigned addresses are not only used
for randomization.  Therefore, after partitioning the corpus, we separate the
locally assigned \ac{MAC} addresses that are used for services such as \ac{P2P}
and WiFi-Extenders from those used as randomized addresses for privacy purposes.
Doing so required us to manually inspect the frame attributes and look for
identifying characteristics.

One prevalent \ac{P2P} service that makes use of locally assigned addresses is
WiFi-Direct.  Fortunately, WiFi-Direct operations contain a WiFi-Direct \ac{IE}
(\texttt{0x506f9a,10}).  Specifically, the following attributes are are observed
with all WiFi-Direct traffic: i) WiFi-Direct \ac{IE} is present, ii) the
observed \ac{OUI} is simply the original \ac{OUI} with the local bit set, and
iii) the \ac{SSID} value, if observed, is set with a prefix of \texttt{DIRECT-}.
Furthermore, manual inspection of the packet capture reveals that these devices
use a single locally assigned \ac{MAC} address for all observed probe request
frames. As these devices are not conducting randomization we remove them from
our dataset.

Similarly, Nintendo devices operating in a \ac{P2P} mode are observed
utilizing a locally assigned address.  Associated frames use a
modified Nintendo \ac{OUI}, one with the local bit set.  Additionally, all
Nintendo \ac{P2P} probe requests contain a unique Vendor Specific \ac{IE},
\texttt{0x00:1F:32}, allowing for an efficient identification and removal from our
dataset.

Lastly, the remainder of our service-based locally assigned addresses were
attributed to WiFi extenders forwarding client probe requests.  These were also
identified as modifying their original \ac{OUI} by setting the local bit.
Commonly observed \acp{OUI}, such as Cisco, D-Link, and Belkin indicated a
likely association to infrastructure devices.  We confirmed our assumptions
through manual packet analysis, which showed: i) the \ac{MAC} address never
changes, ii) each unique device probes for only one \ac{SSID}, and iii) devices
with \ac{WPS} attributes clearly indicate wireless extender models.

Table~\ref{table:local} illustrates that 99.12\% of all locally assigned mac
addresses are randomized addresses, representing $\sim$53\% of our total corpus.
While this may seem like it indicates a large rate of adoption for \ac{MAC}
randomization, these addresses do not directly correlate to the number of
unique devices in our dataset.  While globally unique addresses have a 1-to-1
relationship with individual devices, a device performing randomization has a
1-to-many relationship. It is plausible that a device conducting
randomization may have tens of thousands of addresses over a collection period.
Therefore we posit that much less than $\sim$50\% of devices conduct
randomization.

Our goal, to identify and evaluate potential flaws in currently fielded
randomization policies, requires that we must first answer non-trivial questions
about our real-world dataset.  How many devices were actually performing
randomization?  Which manufacturers and models have implemented randomization in
practice and why?  What operating systems are prevalent? Which randomization policies
are actually used?  

As discussed above, we must first identify distinct \emph{bins} of randomization
within the data.  Table~\ref{table:random} highlights the results of this
analysis.  We completed this analysis by evaluating the following; i) the
\ac{MAC} address prefix (\ac{OUI}, \ac{CID}, random), ii) \ac{WPS} attributes,
iii) 802.11 \ac{IE} derived device signatures, and iv) \ac{mDNS} fingerprinting
techniques \cite{ourpaper2}. Lastly, we confirm our analysis using devices
procured by our team and evaluated in a controlled \ac{RF} environment.  We
provide detailed analysis of our methods, results, and answers to our stated
questions in \S\ref{sec:analysis}.

\section{Analysis}
\label{sec:analysis}


\subsection{Android Randomization}
After removing all of the service-based locally assigned \ac{MAC} addresses
described in \S\ref{sec:methRand}, we aim to separate the remaining $\sim$1.388
million addresses into distinct bins.  First we perform a simple query of our
database where we identify the most common three byte prefixes.  We expect that
the prefixes with the highest occurrences will be the \ac{CID} owned by the
representative devices.  Our findings were surprising: first, the Google owned
\ac{CID} \texttt{DA:A1:19}, was by far the most commonly observed prefix
(52,595), while the second most common prefix \texttt{92:68:C3}, observed 8,691
times was not an \ac{IEEE} allocated \ac{CID}, but rather a Motorola owned
\ac{OUI} with the local bit set.  

The remaining ~177k observed three-byte prefixes, each with total occurrences
ranging from a low of two to a high of seven, show no indication of being a defined
prefix or \ac{CID}.  While we expected to see the Google owned \ac{CID}, we also
expected to see additional \acp{CID} configured by manufacturers to
override the default Google \ac{CID}.  

\subsubsection{\texttt{92:68:C3}}
Investigating the \texttt{92:68:C3} prefix in more detail, we see that devices
using this prefix always transmit granular \ac{WPS} details.  This is helpful as
it lets us easily determine the device model (see \S\ref{sec:related}).  First,
the Motorola Nexus 6 is the only device using this prefix. Using the \ac{WPS}
derived \ac{UUID-E} as a unique identifier, we see that there were 849
individual Motorola Nexus 6 devices in our dataset.  Second, in order to
retrieve the global \ac{MAC} address we use the \ac{UUID-E} reversal technique
previously mentioned \cite{vanhoefRandomMAC, ourpaper2}.  We find that the
actual prefix of the device's \ac{MAC} address is not the expected
\texttt{90:68:C3} \ac{OUI}.  Rather, we observe a set of different Motorola
owned \acp{OUI}.  In combination with with the \emph{config.xml} file (see
Appendix \ref{wpasup}) retrieved from publicly available repositories we
identify that the prefix \texttt{92:68:C3} was purposefully set by Motorola to
replace the Google owned \ac{CID}.

Searching open source Android code repositories revealed no additional
\emph{config.xml} defined prefixes other than the Google and Motorola ones.
This matches what we observe in our real-world dataset.

\subsubsection{\texttt{DA:A1:19}}

The analysis of the Google \ac{CID} \texttt{DA:A1:19} proved more complex,
having serious implications to prior work in derandomization attacks.  Unlike
the Motorola prefix, not all devices using the Google \ac{CID} transmit \ac{WPS}
attributes.  This had multiple effects on our analysis.  First, we were unable
to easily identify the manufacturer and model information when no \ac{WPS}
information was present. Lacking a \ac{UUID-E}, we were unable to precisely
identify total device counts.  More importantly, we were unable to retrieve the
global \ac{MAC} address via the reversal technique. Surprisingly, only
$\sim$19\% of observed \ac{MAC} addresses with the Google \ac{CID} contain
\ac{UUID-E} values.  Since the reversal technique of \citet{vanhoef2} require a
\ac{UUID-E}, this emphasizes the fact that previous evaluations are
insufficient.  A large majority of Android phones are not vulnerable to
\ac{UUID-E} reversal, despite how valuable the technique initially seems.

We evaluated the 8,761 addresses that have \ac{WPS} values before attempting to
breakdown the 43,924 \texttt{DA:A1:19} \ac{MAC} addresses with no \ac{WPS}
information.  We observed a diverse, yet limited spread of manufacturers and
models, depicted in Table~\ref{table:daMan}. Huawei was the most prevalent
manufacturer observed, primarily attributed to the (Google) Nexus 6P (1660
unique devices).  Various versions of the Huawei Mate and Huawei P9 were also
commonly observed.  Sony was well-represented with 277 unique devices across 23
variations of Xperia models.  There were several surprising observations in this
list, namely that Samsung was absent despite having the largest market share for
Android manufacturers \cite{marketshare}.  Blackberry, HTC, and LG were also
poorly represented.  The Blackberry device models were actually four derivations
of the Blackberry Priv, accounting for 277 unique devices observed.  HTC was
largely represented by the HTC Nexus 9 from the Google Nexus line, which
explains the likely use of randomization.  The HTC One M10 was the remaining HTC
device and was only observed once. The only observed LG device was the LG G4
model. We provide a full device breakdown in Appendix~\ref{da_breakdown}.

\begin{table}[!t]
  \centering
  \normalsize
  \caption{\texttt{DA:A1:19} Manufacturer Breakdown}
  \vspace{-2mm}
  \label{table:daMan}
  \scriptsize{
  \begin{tabular}{ l | l | l }\hline 
  \textbf{Manufacturer} & \textbf{Total Devices} & \textbf{Model Diversity} \\ \hline \hline 
  Huawei      & 1708  & 11 \\ \hline
  Sony        & 277   & 23 \\ \hline
  BlackBerry  & 234   & 4  \\ \hline
  HTC         & 108   & 2  \\ \hline
  Google      & 13    & 2  \\ \hline 
  LG          & 1     & 1  \\ \hline 
  \end{tabular}
  }
\end{table}

In all, devices having randomized \ac{MAC} addresses with a Google \ac{CID} and
containing \ac{WPS} attributes amount to a total of 2,341 unique devices.
Taking into account the 849 unique Motorola Nexus 6 devices, only 3,188 devices
spanning 44 unique models are susceptible to the \ac{UUID-E} reversal attack.
Effectively, $\sim$99.98\% of the locally assigned \ac{MAC} addresses in our
corpus are not vulnerable to the \ac{UUID-E} attack.  Furthermore, our corpus
contains approximately 1.2 million client devices with globally unique \ac{MAC}
addresses and over 600 manufacturers and 3,200 distinct models using \ac{WPS}
data fields.  This begs the question, are a large number of Android devices not
conducting randomization?  Do we expect the 43,924 randomized addresses
using the Google \ac{CID} that did not not transmit \ac{WPS} information to make up
all remaining Android devices? 

We attempt to answer these questions by evaluating the 43,924 \texttt{DA:A1:19}
\ac{MAC} addresses where no \ac{WPS} derived data is available.  The process 
proceeds as follows:
\begin{enumerate}
\item Divide
the entire bin into segments, based on the device's signature described in
\S\ref{sec:methRand}, resulting in 67 distinct device signatures, with a starting
hypothesis that each signature represents a distinct model of phone.  
\item For each
signature, parse every packet capture file where that device signature
and the \ac{CID} \texttt{DA:A1:19} were observed.  
\item Apply to our parsing
filter our custom Tshark device signature and limit to probe request frames.
\end{enumerate}

The output of the algorithm is the source \ac{MAC} address, sequence
number, \ac{SSID}, and device signature. 

Left with 2,858 output files, each mapping a device signature with distinct
packet capture, we systematically retrieve the global \ac{MAC} addresses for the
randomized devices.  We will describe in detail the methods for derandomization
for this portion of the dataset in \S\ref{sec:flaws}.  After we obtain the
global \ac{MAC} address for the set of randomized \ac{MAC} addresses within each
bin, we attempt to identify the device model using a variety of techniques.  It
is trivial to identify the manufacturer as the \ac{OUI} provides sufficient
resolution.  However, in order to conjecture the device model we borrow from the
work of \cite{ourpaper2} in which we obtain model granularity from \ac{MAC}
address decomposition.  Next, we look for any case where a device using a global
\ac{MAC} address as the source of a probe request matches the desired signature
and also transmitted a \ac{mDNS} packet at some point.  For this subset we
simply retrieve the model information from the \ac{mDNS} packet
\cite{ourpaper2}.  This leaves us with guesses as to what devices randomize
\ac{MAC} addresses using the \texttt{DA:A1:19} \ac{CID} and transmit no granular
\ac{WPS}-derived model data.  We posit that our set of 67 signature bins can be
condensed into groups of similar signatures based on our derived model
correlations.

In order to better evaluate our assumptions, and now that we have a smaller,
manageable set of possible devices, we procure devices for lab testing. We test
each device using an \ac{RF} enclosed chamber to ensure we limit our collection
to only our individual test phones.  We leave each device in the chamber for
approximately five minutes, collecting only the probe requests.

We evaluate the collection results by comparing to our derived signatures and
ask the following: do we observe \ac{MAC} address randomization? If so, does the
device signature match expectations when using a global address? Similarly, does
the device signature match expectations when using a randomized address?  Our
findings are presented in Table~\ref{table:noWPSdevices}.

\begin{table}[!t]
  \centering
  \caption{\texttt{DA:A1:19} no \ac{WPS}}
  \vspace{-2mm}
  \label{table:noWPSdevices}
  \scriptsize{
  \begin{tabular}{l| c| c} \hline 
    \textbf{Category} & \textbf{Confirmed} &  \textbf{\% of no WPS}  \\ \hline \hline 

    \texttt{Bin 1}            &            & 57.7\%   \\ \hline
    \ti \texttt{LG Nexus 5X}  & $\surd$    &          \\ \hline
    \ti \texttt{Google Pixel} & $\surd$    &          \\ \hline

    \texttt{Bin 2}            &            & 18.5\%   \\ \hline
    \ti \texttt{LG G5}        & $\surd$    &          \\ \hline
    \ti \texttt{LG G4}        & $\surd$    &          \\ \hline

    \texttt{Bin 3}            &            & 2.0\%   \\ \hline
    \ti \texttt{OnePlus 3}    & $\surd$    &          \\ \hline
    \ti \texttt{Xiaomi Mi Note Pro}  & $\surd$   &          \\ \hline

    \texttt{Bin 4}            &            & .2\%   \\ \hline
    \ti \texttt{Huawei}       &  $\surd$   &          \\ \hline
    \ti \texttt{Sony}         &  $\surd$   &          \\ \hline

    \texttt{Bin 5}            &            & 2.6\%   \\ \hline
    \ti \texttt{Cat S60}      &  $\surd$   &          \\ \hline

    \texttt{Bin 6}            &            & 12.2\%   \\ \hline
    \ti \texttt{Composite}    &  $\surd$   &          \\ \hline

    \texttt{Bin 7}            &            & 6.8\%   \\ \hline
    \ti \texttt{Unknown}      &            &          \\ \hline

  \end{tabular}
  }
\end{table}

Bin 1 is represented by the Google devices LG Nexus 5X and Google Pixel.  This
bin encompasses 57.7\% of the 43,924 \ac{MAC} addresses observed using the
Google \ac{CID} without \ac{WPS} data.  It is prudent to mention that we cannot
claim that is an exhaustive list of devices implementing randomization using this set of
signatures. 

Next, we evaluate bin 2, representing 18.5\% of the category's total.  We
observe only LG devices, specifically we posit that LG G series devices make up
this subset.  We confirm that both the LG G4 and G5 devices match the signatures
and behavior of this bin.  We surmise that additional G series devices are
represented, however we have no validation at this time.  Worth mentioning is
that the LG G4 and Pixel identified in the previous \texttt{DA:A1:19} with
\ac{WPS} section were only observed because a \ac{WPS} action was triggered.  By
default, \ac{WPS} data is not transmitted by the devices in our \emph{no-wps}
category.  We confirm this analysis in our lab environment, observing \ac{WPS}
data fields only when the user triggers a \ac{WPS} event.

In bin 3, a smaller bin (2\%), the OnePlus 3, and the Xiamoi Mi Note Pro
are representative of the identified signatures.  

Bin 4, the smallest of our bins with less then one percent of our dataset,
consisted of Huawei and Sony devices.  These are devices seen using
\ac{WPS}, but in some frames do not include the \ac{WPS} data fields.

The Cat S60 smartphone was the only device identified in bin 5.  As in other
bins, we make no claim that no other devices share this signature. 

Bin 6 represents a combination of the aforementioned devices observed in the
various bins. This is caused by a device, that on occasion rotate between a
standard device signature and a stripped down version with limited 802.11
\ac{IE} fields.  An example of this signature behavior is described in
\S\ref{sec:dev_sig} and depicted in Figure~\ref{fig:sig}. As such, this bin is
represented by the previously mentioned devices.

We fail to identify anything with any sense of confidence within bin 7.

\subsubsection{Motorola}

After an exhaustive look at the randomization schemes employed by Android we
still lack any evidence of \ac{MAC} address randomization by Samsung or Motorola
devices (other then the Google based Motorola Nexus 6).  We attempt to find any
evidence of non-standard randomization employed by these models by looking at
probe requests with globally assigned \ac{MAC} addresses.  In a similar manner
to how we identified the most common prefixes for locally assigned addresses, we
attempt to identify \acp{OUI} with unusually high occurrences within individual
packet captures.  Our premise is that this will indicate the use of an \ac{OUI}
as a prefix for a set of randomized \ac{MAC} addresses.

We first ruled out all \ac{P2P} service related addresses as previously
described, leaving a single manufacturer of interest - Motorola.  We identified
multiple occurrences of various Motorola \acp{OUI} with an abnormally high
percentage of the unique addresses in a packet capture.  After inspecting forty
captures with this anomaly we confirmed that a subset of Motorola devices
perform randomization using neither a \ac{CID} nor an \ac{OUI} with the
local bit set.  These devices used one of several Motorola owned
\acp{OUI}, using the global \ac{MAC} address occasionally, and a new randomized
\ac{MAC} address when transmitting probe requests.

This is an especially strange result because it shows that Motorola is using
randomized global addresses.  This violates the core expectation that no two
devices will use the same global \ac{MAC} address.  In particular, it is
possible for one of these devices to temporarily use the true, global \ac{MAC}
address of another device as one of its random addresses.

We identified two distinct signatures consistently observed within this Motorola
dataset.  Using the aforementioned \ac{mDNS} techniques to guess a device model
we posit that one signature belongs to the Moto G4 model while the second
corresponds to a Moto E2.  We acquired Moto G4 and E2 smartphones and confirmed
our hypothesis. Additionally, we observed that a Moto Z2 Play device model shares
the same randomization behavior and signature as the Moto G4.

\subsubsection{Samsung}

It is interesting to note that we never observed Samsung devices performing
\ac{MAC} address randomization, despite being the leading manufacturer of
Android smartphones.  Samsung uses their own 802.11 chipsets, so it is possible
that chipset compatibility issues prevent implementing randomized \acp{MAC}
addresses. Samsung devices alone represent $\sim$23\% of Android devices in our data
set, contributing substantially to the low adoption rate that we see. 

\subsection{iOS Randomization}

After completing the randomization analysis of Android devices, we still have
over 1.3 million \ac{MAC} addresses not attributed to any randomization scheme.
Next we turn to the analysis of iOS randomization.

Upon the release of iOS 8.0, Apple introduced \ac{MAC} address randomization,
continuing with minor but valuable updates to the policy across subsequent iOS
releases.  We were faced with an immediate dilemma, how do we identify iOS
associated probe requests? Apple iOS devices do not transmit \ac{WPS} fields to
indicate any sort of model information, and we had no knowledge of any Apple
owned \ac{CID}.  In order to identify any prefix pattern we once again
utilized our \ac{RF}-clean environment to test Apple device behavior.  Our goal
was to create as many randomized \ac{MAC} addresses as possible from a device
and look for a pattern in the resulting prefixes.  To force a new randomized \ac{MAC}
address we simply enable and disable WiFi mode repeatedly.  

Our initial thought was that Apple would use a \ac{OUI} or \ac{CID} like other
manufacturers and simply randomize the least significant 24 bits of the \ac{MAC}
address.  However, we quickly found that the \ac{MAC} addresses randomly
generated by iOS devices do not share any common prefix.  In fact, they appear
to be completely random, including the 24 \ac{OUI} bits, except for the local
bit which is always set to 1 and the multicast bit which is set to 0.  To lend
credence to this new hypothesis we sampled 47,255 random \ac{MAC} addresses from
an iOS device and ran standard statistical tests to determine if they were
uniformly distributed (see Appendix \ref{randtests}).  These tests confirmed
that, with the exception of the local and unicast bit, iOS most likely
implements true randomization across the entire \ac{MAC} address. This is
interesting given the fact that the \ac{IEEE} licenses \ac{CID} prefixes for a
price, meaning that Apple is freely making use of address space that other
companies have paid for.

Based on these findings, we are faced with identifying a randomization scheme
where randomness is applied across $2^{46}$ bits of the byte structure.  We can
not simply assume that if the prefix does not match an offset of an allocated
\ac{OUI} that it is an iOS device.  This is due to the aforementioned clobbering
of other manufacturers \ac{OUI} space.  Our next step was to leverage the use of
\ac{mDNS} once again.  We take the union of global \ac{MAC} addresses derived
from probe requests that are also seen as source addresses for iOS related
\ac{mDNS} packets.  This results in a set of probe requests that we can confirm
are Apple iOS devices.  We then extract all of the signatures for these devices.
We suspected that this retrieved only a portion of the relevant iOS signatures.
Next we collected signatures from all of our Apple iOS lab test devices using
our \ac{RF} enclosure.  Finally, we identify signatures of all remaining locally
assigned \ac{MAC} addresses in which we have no assigned categorization.  We then
seek to find any probe requests with global source address that have matching
signatures.  If the \ac{OUI} of the global addresses resolves to an Apple
\ac{OUI} we consider that a valid signature.  This is slightly different then
our \ac{mDNS} test as we cannot attribute the signature to a specific set of iOS
device models.  We test our entire iOS signature set and ensure that no non-iOS
global \ac{MAC} addresses are ever observed with these signatures.  

In June 2016, midway through our research, iOS 10 was released. Inexplicably the
addition of an Apple vendor specific \ac{IE} was added to all transmitted probe
requests.  This made identification of iOS 10 Apple devices trivial regardless
of the use of \ac{MAC} address randomization.  We believe the difficulty of
identifying \ac{MAC} address randomization to be one of the best countermeasures
to defeating randomization. Compounding our incredulity, the data field
associated with this \ac{IE} never changes across devices. 

Using our combined  set of all Apple iOS signatures, we identify $\sim$1.3 million
distinct randomized \ac{MAC} addresses, by far the most populous (94.7\%) of our
randomization categories.

\subsection{Windows 10 and Linux Randomization}

To conclude our categorization of randomization schemes, we look to identify the
probe requests from devices using Windows 10 and Linux \ac{MAC} address
randomization implementations. Our first test compares the signatures obtained
from laboratory laptops to the signatures of our locally assigned dataset.  We
find 59 matches to our laptop signatures, indicating possible Windows 10 or
Linux randomization.  Next, we parse collection files using the locally assigned
\ac{MAC} addresses from the probe request frames of these devices.  Our
hypothesis, if we find matching locally assigned \ac{MAC} addresses in
authentication, association, or data frames, that the randomizations scheme is
likely Windows 10 or Linux.  This assumption is due to the fact that the
randomization policies use the same locally assigned address for network
establishment and higher layer data frames. To that end, we find that 14 of the
59 devices assessed to be Windows/Linux computers use a locally assigned
\ac{MAC} address when associated to a network.

\section{\ac{MAC} Randomization Flaws}
\label{sec:flaws}

Now that we have a baseline understanding of the randomization implementations
used by modern mobile \acp{OS} we are able to assess for vulnerabilities.

\subsection{Adoption Rate}

The most glaring observation, while not necessarily a flaw per se, is that the
overwhelming majority of Android devices are not implementing the available
randomization capabilities built into the Android \ac{OS}.  We expect that this
may be partly due to 802.11 chipset and firmware incompatibilities.  However,
some non-randomizing devices share the same chipsets as those implementing
randomization, so it is not entirely clear why they are not utilizing
randomization. Clearly, no effort by an attacker is required to target these
devices.   

\subsection{Global Probe Request}

We next explore the flaws of the observed \ac{MAC} address randomization
schemes.  One such flaw, the inexplicable transmission of the global \ac{MAC}
address in tandem with the use of randomized \ac{MAC} addresses.  We observe
this flaw across the gamut of Android devices.  The single device in which we do
not observe this was the Cat S60 smartphone.  In no instance did the Cat S60
transmit a global \ac{MAC} address probe request, except immediately prior to an
association attempt. Exploiting this flaw it was trivial to link the global and
randomized \ac{MAC} addresses using our device signatures and sequence number
analysis.  Between probe requests, the sequence numbers increase predictably so
an entire series of random addresses can be linked with a global address by just
following the chain of sequence numbers.  While using sequence numbers has been
discussed before in prior work \cite{vanhoefRandomMAC}, the fact that the global
\ac{MAC} address is utilized while in a supposedly randomized scan state has
not.  This strange behavior is a substantial flaw, and effectively negates any
privacy benefits obtained from randomization. 
In our lab environment we observed that in addition to periodic global \ac{MAC}
addressed probe requests, we were able to force the transmission of additional
such probes for all Android devices.  First, anytime the user simply turned on
the screen, a set of global probe requests were transmitted.  An active user, in
effect, renders randomization moot, eliminating the privacy countermeasure all
together.  Second, if the phone received a call, regardless of whether the user
answers the call, global probe requests are transmitted.  While it may not
always be practical for an attacker to actively stimulate the phone in this
manner, it is unfortunate and disconcerting that device activity unrelated to
WiFi causes unexpected consequences for user privacy.

\subsection{\ac{UUID-E} Reversal}

Vanhoef et. al. introduce the \ac{UUID-E} reversal attack against Android
devices \cite{vanhoefRandomMAC}.  Devices transmitting probe request frames with
\ac{WPS} enriched data fields, specifically, the \ac{UUID-E} are vulnerable to a
reversal attack where the global \ac{MAC} address can be retrieved using the
\ac{WPS} \ac{UUID-E} value.  The flaw caused by the construction of the
\ac{UUID-E}, where the \ac{MAC} address is used as an input variable along with
a non-random hard-coded seed value.  This implementation design flaw allows for
the computation of pre-computed hash tables, whereby retrieving the global
\ac{MAC} address requires only a simple search of the hash tables.  This
revelation, both groundbreaking and disconcerting, still leaves the reader to
guess as to the plausibility of the attack against randomized devices.  We find
several issues with their approach, specifically in respect to derandomization
analysis: i) randomization was not employed in 2013, when the data used in their
evaluation was gathered ii) anonymized data eliminates accuracy checks, and iii)
removing locally assigned \ac{MAC} addresses effectively eliminates the ability
to evaluate the attack against devices performing randomization.

Accordingly, we use our corpus of \texttt{DA:A1:19} and \texttt{92:68:C3}
datasets to evaluate the effectiveness and viability of the \ac{UUID-E} attack.
Our foremost observation is that only 29\% of random \ac{MAC} addresses from
Android devices include \ac{WPS} attributes.  Effectively 71\% of  this Android
dataset is completely immune to the \ac{UUID-E} reversal attack.  This is in
addition to the fact that iOS devices are wholly immune to the attack, as they
do not use \ac{WPS}.   We refer back to Table~\ref{table:daMan} the limited
number of Android models performing randomization and transmitting the necessary
\ac{WPS} \ac{UUID-E} attribute.

We then retrieve the global \ac{MAC} address from the probe requests of these
devices that used both random and global \ac{MAC} addresses, exploiting the
previously discussed flaw.  We use this set of 1,417 \emph{ground truth}
\ac{MAC} addresses to test the effectiveness of the \ac{UUID-E} reversal attack.
First we pre-compute the required hash tables.  To build hash tables for the
entire \ac{IEEE} space would be non-trivial, requiring significant disk space
and processing time.  While an exhaustive compilation of the address space is
certainly possible, we use the knowledge gained from decomposing the
randomization schemes to efficiently construct our tables. We build the hash
tables using only the \acp{OUI} owned by manufacturers we have observed to
implement randomization.  The resulting hash table is a manageable 2.5TBs, where
using pre-sorting techniques, we can retrieve an \ac{UUID-E}'s global \ac{MAC}
address in $<$ 1 second.

We retrieve a global \ac{MAC} address for 3,187 of the 3,188 \acp{UUID-E}. In
previous work it was left inconclusive whether the retrieved \ac{MAC} addresses
were in fact the global 802.11 \ac{MAC} address or instead the Bluetooth
\ac{MAC} address. The \ac{UUID-E} derived from the HTC One M10 device, was the
example \ac{UUID-E} listed in the \emph{wpa\_supplicant.conf} file.  With exception
of the HTC Nexus 9, all HTC phones in our dataset (regardless of randomization)
used this non unique \ac{UUID-E}.

Comparing the 1,417 \emph{ground truth} addresses to those retrieved from the
\ac{UUID-E} attack we achieve a 100\% success rate.  Indicating that the
retrieved addresses are in fact the global 802.11 \ac{MAC} addresses, completing
the missing link from the evaluation of \citet{vanhoefRandomMAC}.

\begin{figure*}[t!]
\begin{align*}
  \text{$Sig_{G}$}  &= 
  \texttt{0,1,50,3,45,221(0x50f2,8),htcap:012c,htagg:03,htmcs:000000ff} \\
  \text{$Sig_{R}$}  &= 
  \texttt{0,1,50} \\
\end{align*}

  \vspace{-12mm}
  \caption{Device Signature (Motorola Moto E2)}
  \label{fig:sig}
\end{figure*}

\subsection{Device Signature}
\label{sec:dev_sig}

To aide in derandomization we employ fingerprinting techniques, using signatures
derived from the 802.11 \acp{IE} borrowed from previous work
\cite{vanhoefRandomMAC, googlepaper}.  We used this technique first to aide in
the identification of the randomization schemes employed by Android and iOS
devices.  

This technique allows us to remove all extraneous probe request traffic,
providing us a ``cleaner'' dataset in which to employ sequence number analysis.
We modify the Wireshark files \emph{packet-ieee80211.c} and
\emph{packet-ieee80211.h}, creating a new dissector filter,
\emph{device.signature}.  We are able to filter previous collection files as
well as conduct filtering on live collection.  While our contribution to the
Wireshark distribution is novel, the fingerprinting technique is not, as we
borrowed from related work.  However, prior work tested against datasets not
performing randomization which fails to provide accurate context.  We test the
signature technique against our real world corpus, revealing flaws in previous
signature based attacks.

Regardless of the Android implementation, a device transmits probe request
frames which have varying signatures (based on \acp{IE}, see
\S\ref{sec:related}).  Devices often use two or more signatures while using a
global \ac{MAC} address, so simply using the signature is insufficient.
Additionally, the same holds for randomized addresses, in which we observe
multiple signatures.  In both cases, the second signature, has minimal 802.11
\acp{IE}.  Due to the fact that nearly all devices periodically use this
signature, it creates significant complexity to any signature based
derandomization attack.  Finally, as Figure~\ref{fig:sig} illustrates, we
observe that most Android devices use different signatures when randomizing
compared to when using a global \ac{MAC} address.  As such, previously described
signature-based tracking methods fail to correlate the addresses.  Using our
decomposition of Android randomization schemes, and the derived knowledge of how
distinct bins of devices behave,  we properly pair the signatures of probe
requests using global and randomized \ac{MAC} addresses. Only by combining these
signatures are we able to accurately and efficiently retrieve the global
\ac{MAC} address.

We observe no such change in signatures of iOS devices within a collection
timeframe. While an iOS device may not use alternate signatures, they do not
send globally addressed probe requests.  Therefore, at this juncture, we have
not identified a method of resolving the global \ac{MAC} address.

\subsection{Association/Authentication Frames}

We observe that Android and iOS devices use sequential sequence numbers across
management frame types.  Using only passive analysis we can follow a devices
transition from randomized probe requests to an authentication or association
frame by following the sequence numbers. This is particularly useful as all
authentication and association frames from iOS and Android devices use the
global \ac{MAC} address.  Using the techniques described in \cite{googlepaper}
we create a set of signatures for the association frames of iOS devices,
specifically to aide in confirmation that the device observed in the probe
request is also the same device type as the association frame.  This method
relies on the targeted device attempting to establish a network connection with
a nearby \ac{AP}.  As this is fairly user-activity dependent, we reinvestigate
the plausibility of the Karma attack against current randomization schemes. 

\subsection{Karma Attack} The current versions of iOS and Android randomization
policies have eliminated the vast majority of cases where a \emph{directed probe}
is used.  A directed probe is a probe request containing a specified \ac{SSID}
that the device wishes to establish a connection (a previously known or
configured \ac{SSID}), as opposed to a broadcast probe which solicits a response
from all \acp{AP} in range.  Today, the predominant use of broadcast probes has
directly effected the ability for a Karma-based attack to succeed.  Karma-based
attacks work by simulating an access point that a device prefers to connect to.
A variety of implications such as man-in-the-middle attacks are common follow-on
consequences, however we are only interested in retrieving the global \ac{MAC}
address and therefore require only a single authentication frame to be
transmitted by a targeted device. To this end Vanhoef et. al. also investigate
Karma attacks, implemented via a predefined top-n \ac{SSID} attack, achieving a
17.4\% success rate, albeit not specifically related to devices performing
randomization.

Unlike previous work, we observe devices while in a randomized state in order to
identify specific behaviors that directly counteract randomization privacy
goals.  Specifically, do we observe traits that allow for a targeted Karma
attack?  It is well known that hidden networks require directed probes, so while
this is a vulnerability to randomization, it is fairly uncommon, and a decision
in which a user chooses to implement.  Similarly, previous connections to ad hoc
networks, saved to the devices network list, cause both Android and iOS devices
to send directed probes.  As with hidden networks, this uncommon condition
requires action from the user, however when observed, the Karma attack is viable. 

Finally, we observe a more disconcerting trend: devices configured for seamless
cellular to WiFi data-offloading, such as Hotspot 2.0, EAP-SIM and EAP-AKA force
the use of directed probes and are inherently vulnerable to Karma-based attacks
\cite{skycure}.  The expanding growth of such handover polices reveals a
significant vulnerability to randomization countermeasures.  Further
exasperating the problem, these devices are pre-configured with these settings,
requiring no user interaction.  We confirmed these settings by inspecting the
\emph{wpa\_supplicant.conf} file of a Motorola Nexus 6 and Nexus 5X.  Removing the
networks from the configuration file requires deletion by a rare user with both
command line savvy and awareness of this issue.

We test for the presence of these network configurations in our corpus by
evaluating all randomized addresses using \ac{WPS} fields.  We are able to
accurately evaluate unique devices using the \ac{UUID-E} value as the unique
identifier.  We filter for any instance where the device sends a directed probe,
retrieving the \ac{SSID} value for each.  Sorting by most common occurrence the
top three most common \acp{SSID} were \texttt{BELL\_WIFI}, \texttt{5099251212},
and \texttt{attwifibn}. The \acp{SSID} \texttt{BELL\_WIFI} and
\texttt{5099251212} are used by the mobile carrier Bell Canada for seamless WiFi
offloading.  Interestingly, the \texttt{attwifibn} \ac{SSID} is related to free
WiFi hotspots provided by the Barnes and Noble bookstore.   Only $\sim$5\% of
the datasets 3,188 devices transmitted a directed probe.  However, of those that
did,  17\% of were caused by the preconfigured mobile provider settings.

Next we take a cursory look at Apple iOS and Android devices with no amplifying
\ac{WPS} information.  We do not get precise statistics, however, we observe the
same trend. 


\subsection{Control Frame Attack}
We now evaluate active attack methods for identifying a device by its global
\ac{MAC} address while in a randomized state.  Our premise: can we force a
device performing \ac{MAC} address randomization to respond to frames targeting
the global \ac{MAC} address?  This would allow for easy tracking of devices,
even when they are randomizing, because an active attacker could elicit a
specific response from them at any time if they are within wireless range.

\begin{table}[h]
  \centering
  \caption{Class 1 Frames \cite{opac-b1128195}}
  \vspace{-2mm}
  \label{table:frames}
  \scriptsize{
  \begin{tabular}{ l | l | l }\hline 
  \textbf{Control} & \textbf{Management} & \textbf{Data} \\ \hline \hline 
    RTS        & Probe Request   & Frame w/DS bits false \\ \hline
    CTS        & Probe Response  &  \\ \hline
    Ack  & Beacon          &  \\ \hline
    CF-End          & Authentication  &  \\ \hline
    CF-End+CF-Ack   & Deauthentication&  \\ \hline
                    & ATIM            &  \\ \hline

  \end{tabular}
  }
\end{table}

\begin{figure}[t]
\begin{tikzpicture} [->, node distance=3cm, auto,every node/.style={align=center}]
\tikzstyle{state} = [rectangle, rounded corners, draw=black, very thick, text width=10em, text centered]
\tikzstyle{line} = [draw, thick, <-]
  \node [state] (state3) {{\bf State 3} \\ Authenticated and associated};
  \node [state, below of=state3] (state2) {{\bf State 2} \\ Authenticated and unassociated};
  \node [state, below of=state2] (state1) {{\bf State 1} \\ Unauthenticated and unassociated};
  \path (state3) edge [loop left, thick,out=195, in=170, distance=.8cm] node {Class 1, 2 \\ and 3 frames} (state3);
   \path (state2) edge [loop left, thick,out=195, in=170, distance=.8cm] node {Class 1 and 2\\ frames}  (state2);
    \path (state1) edge [loop left, thick,out=195, in=170, distance=.8cm] node {Class 1 frames} (state1);
  \path ([xshift=-2ex]state3.south) edge [thick] node [anchor=east] {Disassociation} ([xshift=-2ex]state2.north);
  \path ([xshift=-2ex]state2.south) edge [thick] node [anchor=east]{Deauthentication} ([xshift=-2ex]state1.north);
    \path ([xshift=2ex]state2.north) edge [thick] node [anchor=west] {Association} ([xshift=2ex]state3.south);
  \path ([xshift=2ex]state1.north) edge [thick] node [anchor=west]{Authentication} ([xshift=2ex]state2.south);
 \end{tikzpicture}
\caption{802.11 State Diagram}
  \label{fig:stateD}
\end{figure}
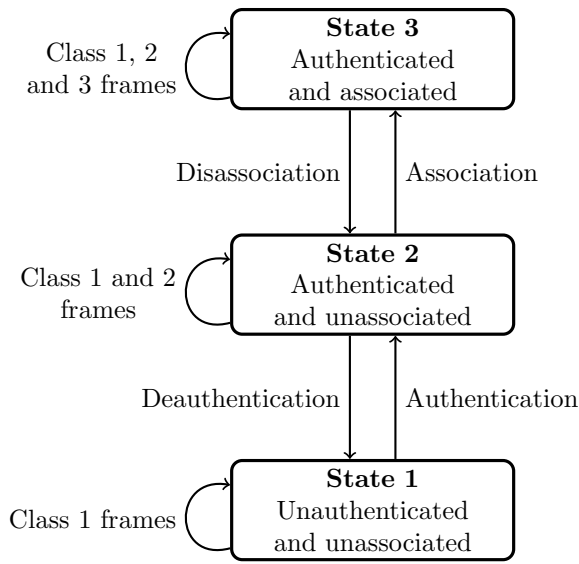

Figure~\ref{fig:stateD} depicts the 802.11 state diagram illustrating the
various states of association for 802.11 devices \cite{opac-b1128195}.  We are
particularly interested in the frame types that can be sent or received while in
an unauthenticated and unassociated state (State 1).  The frame types (Class 1
frames) allowed while in State 1 are depicted in Table~\ref{table:frames}.

In our lab environment, we use packet crafting tools (SCAPY, libtins) to
transmit customized packets for each frame type, targeting the global \ac{MAC}
of the device. 

The source \ac{MAC} address of the frame is a uniquely crafted \ac{MAC} address.
It is not the actual \ac{MAC} address of our transmitter.  This ensures that we
can accurately track any responses to our crafted message, removing any possible
control frames that happen to be sent to the actual transmitter address. Of the
twelve Class 1 frame types used for the attack, we successfully elicited a
response from only the \ac{RTS} frame.  

Request to Send and Clear to Send (\ac{RTS}/\ac{CTS}) transmissions are
available in the \ac{IEEE} 802.11 specification as part of a Carrier Sense
Multiple Access with Collision Avoidance scheme.  When a node desires to send
data an \ac{RTS} may be sent to alert other nodes on the channel that a
transmission is about to begin and the period of time during which they should
not transmit on that channel so as to avoid collisions.  If there are no
conflicting uses of the channel, the target node will respond with a \ac{CTS} to
acknowledge the request and give the transmitting node permission to solely
communicate on the medium.  

As for previous location and tracking attacks, some researchers have used
\ac{RTS}/\ac{CTS} messages to perform Time of Arrival computations \cite{Hoene}
while others have extended these techniques to perform Time Difference of
Arrival calculations from timestamps in exchanged frames \cite{Cui}.  These
older methods perform localization on Access Points from client devices.  The
novelty in our method is that we are sending \ac{RTS} frames to IEEE 802.11
client devices, not \acp{AP}, to extract a \ac{CTS} response message which we
derive the true global \ac{MAC} address of that device.  Instead of a
localization attack, we are using \ac{RTS}/\ac{CTS} exchanges to perform
derandomization attacks.

The result of sending a \ac{RTS} frame to the global \ac{MAC} address of a
device performing randomization was that the target device responded with a
\ac{CTS} frame.  A \ac{CTS} frame, having no source \ac{MAC} address, is
confirmed as a response to our attack based on the fact that it was sent to 
the original, crafted source \ac{MAC} address.  A full device listing utilized
for the control frame attack is available in Appendix~\ref{rts_breakdown}.

Once the global \ac{MAC} address is known, that device can be easily tracked
just as if randomization were never enabled.  This might cause one to wonder why
vendors would go to such lengths to include \ac{MAC} address randomization in a
device only to allow that same device to divulge the protected information
through an administrative protocol.  We assert that this phenomenon is beyond
the control of individual vendors.  The fact is that this behavior occurs across
the board on every device we have physically tested as shown in
Appendix~\ref{rts_breakdown}.   This leads us to believe that
\ac{RTS}/\ac{CTS} responses are not a function of the OS, but of the underlying
\ac{IEEE} 802.11 chipset.  Manufacturers have configured their chipset hardware
with default \ac{RTS}/\ac{CTS} operation which may not even be accessible to
configure at the \ac{OS} level.  If we are correct, this derandomization issue
can not be fixed with a simple patch or OS update.  Susceptible mobile devices
will be unmasked by this method for the lifetime of the device.  Additionally,
due to the hardware level nature of this phenomenon, there will be a significant
delay in the market until mobile devices resistant to this attack are produced,
assuming manufacturers recognize this as a flaw and subsequently design a
process truly capable of delivering MAC address privacy.

\begin{table*}[!t]
  \centering
  \caption{Derandomization Technique Results}
  \vspace{-2mm}
  \label{table:derando}
  \scriptsize{
  \begin{tabular}{*{6}{c}} \hline 
  \textbf{Randomization Bin} & \textbf{\ac{UUID-E} Reversal} & \textbf{Global
    \ac{MAC} Address} 
    & \textbf{Auth/Assoc} & \textbf{Hotspot 2.0 - Directed Probes} &
    \textbf{RTS Attack}  \\ 
    &&Probe Request& Frames& \textbf{Karma Attack} & \\ \hline \hline

    \texttt{DA:A1:19} with \ac{WPS}     & $\surd$  & $\surd$ & $\surd$  & $\surd$  & $\surd$   \\ \hline
    \texttt{DA:A1:19} w/o \ac{WPS}      & $\times$ & $\surd$  & $\surd$ &$\surd$  & $\surd$   \\ \hline
    \texttt{92:68:C3} with \ac{WPS}     & $\surd$  & $\surd$  & $\surd$ & $\surd$  & $\surd$   \\ \hline
     Motorola (No local bit)        & $\times$ & $\times$ & $\surd$ & $\surd$  & $\surd$   \\ \hline 
     Apple iOS                          & $\times$ & $\times$ & $\surd$ & $\surd$  & $\surd$   \\ \hline 

  \end{tabular}
  }
\end{table*}

There are multiple scenarios in which a motivated attacker could use this method
to violate the privacy of an unsuspecting user.  If the global \ac{MAC} address
for a user is ever known, it can then be added to a database for future
tracking.  This global MAC address can be divulged using the techniques
discussed in this paper, but it can also be observed any time the user is
legitimately using that global MAC address, such as when connected to an \ac{AP}
at home or work.  This single leakage of the true identifier will allow an
attacker to send an \ac{RTS} frame containing that global MAC address in the
future to which that host will respond with a correct \ac{CTS} when it is in
range.  Conceivably, an adversary with a sufficiently large database and
advanced transmission capabilities could render randomization protections moot.
Additional tests, while the target device had WiFi or Airplane-modes, enabled or
disabled respectively, revealed further concerns.  Namely, Android devices
performing location-service enabled functions wake the 802.11 radio.  Our
\ac{RTS} attack was thusly able to trigger a \ac{CTS} response from the target,
circumventing even extreme privacy countermeasures.

Lastly, we add improvements, using our Wireshark signature filters, to eliminate
the constant barrage of transmitted \ac{RTS} frames.  Our collection algorithm
is pre-loaded with the target of interest's device signature, where upon
observing the signature in the target area we launch the preconfigured \ac{MAC}
address.  We test this against our diverse test phones with 100\% success.

\subsubsection{Bluetooth Correlation}

We offer an additional method to derive the global WiFi \ac{MAC} address for
later use in a \ac{RTS} attack.  \citet{hacking} claim that Apple iPhone
devices, beginning with the iPhone 3G, utilize a one-off scheme for the
allocation of the Bluetooth and WiFi \ac{MAC} addresses, where the \ac{MAC}
address is actually equal to the Bluetooth address, plus or minus one.  Using a
novel algorithm to calculate the WiFi and Bluetooth \ac{MAC} address from iOS
devices operating in hotspot mode, we provide evidence countering this claim.

We identified that Apple iOS devices, operating in hotspot mode, send beacon
management frames containing an Apple vendor specific \ac{IE}.  This \emph{Type
6} field closely resembles the source \ac{MAC} address of the device.  As
Wireshark does not process this field correctly we built custom dissectors to
create display filters for the Apple vendor tag \ac{IE} and associated data
fields.  We first test on 29 Apple iOS lab devices, placing each in hotspot mode
and collecting the beacon frames.  We retrieve the true Bluetooth and WiFi
\ac{MAC} addresses from the device settings menu of the phone.  We then parse
the beacon frames, outputting the  source \ac{MAC} address and six byte Type 6
\ac{IE}. 
 
We observe that the Type 6 field exactly represents the Bluetooth \ac{MAC}
address.  The source \ac{MAC} address of the Beacon frame has the local bit
set.  However, the first byte of the source \ac{MAC} address is not a simple
offset of the global \ac{MAC} address as seen in most \ac{P2P} operations.  To
resolve the actual global \ac{MAC} address we find that replacing the first byte
of the source \ac{MAC} address with the first byte of the Type 6 (Bluetooth
Derived) \ac{MAC} address, we obtain the correct WiFi \ac{MAC} address of the
device.  This permutation is successfully tested for all 29 test devices across
the gamut of model and iOS versions.

Interestingly, six of the 29 test devices did not show a one-off \ac{MAC}
address allocation.  As such, we seek to identify the accuracy of the previous
claim that iOS devices use this one-off scheme by evaluating across our
entire corpus.

A total of 3,576 devices were identified in our dataset containing the Type 6
field of which $\sim$95.4\% utilized a one-off addressing scheme.
Interestingly, $\sim$88.2\% of those devices had a Bluetooth address that was
one-higher then the WiFi \ac{MAC} address.  Indicating that even when the offset
is used it is not uniformly implemented.  We are unsure as to why $\sim$4.6\% of
iOS devices do not use the one-off policy.  Regardless, in all cases the
\ac{OUI} of the two interfaces are the same.  Using the \ac{mDNS} model
correlation analysis we observed no indication that offset scheme is correlated
with the device model.  

\section{Conclusions}
\label{sec:conclusions}

We provide a detailed breakdown of the randomization polices implemented, the
associated device models, and the identification methods thereof.  This
granularly detailed decomposition allowed for fine-tuned improvements to prior
attempts at \ac{MAC} address derandomization as well as providing novel
additions.

Our analysis illustrates that  \ac{MAC} address randomization policies are neither
universally implemented nor effective at eliminating privacy concerns.
Table~\ref{table:derando} depicts the diversity of presented attacks, across
the spectra of randomization schemes and \acp{OS}, highlighted by the \ac{RTS} control
frame attack targeting a widespread low-level chipset vulnerability.

To be truly effective, randomization should be universally adopted.  A continued
lack of adoption, allowing for simpler identification, effectively reduces the
problem set for an attacker.  The more devices performing randomization
within a test set, the harder it will be to diffuse each device's associated
traffic.  This is particularly true if we can continue to bin the various
schemes, further reducing the problem set.

We propose the following best practices for \ac{MAC} address randomization.
Firstly, mandate a universal randomization policy to be used across the spectra
of 802.11 client devices.  We have illustrated that when vendors implement
unique \ac{MAC} address randomization schemes it becomes easier to identify and
track those devices.  A universal policy must include at minimum, rules for
randomized \ac{MAC} address byte structure, 802.11 \ac{IE} usage, and sequence
number behavior.

To reiterate, these best practices can only be truly effective when enforced
across the spectrum of devices. Granular examples of such policy rules:

\bi

\item Randomize across the entire address, providing $2^{46}$ bits of
  randomization. 
\item Use a random address for every probe request frame.
\item Remove sequence numbers from probe requests.  
\item If sequence numbers are used, reset sequence number when transmitting
  authentication and association frames.
\item Never send probe requests using a global \ac{MAC} address.
\item Enforce a policy requiring a minimal and standard set of vendor \acp{IE}.  Move any lost functionality to the
  authentication/association process, or upon network establishment utilize discovery
  protocols.
\item Specifically, the use of \ac{WPS} attributes should be removed except when
  performing \ac{P2P} operations.  Prohibit unique vendor tags such as
  those introduced by Apple iOS 10.
\item Eliminate the use of directed probe requests for cellular offloading.
\item Mandate that chipset firmware remove behavior where \ac{RTS} frames
  received while in State 1 elicit a \ac{CTS} response.   
\ei

\section*{Acknowledgments}
We thank Rob Beverly, Adam Aviv, and Dan Roche for early feedback.


 \bibliographystyle{ieeetr}
  \bibliography{refs}

\clearpage
\appendix

\section{OS Randomization Configuration}
\label{wpasup}

\subsection{Android}

In October 2014 the \emph{wpa\_suppplicant.conf} file, used by Android, Linux,
Windows, and OS X client stations \cite{linuxwpasupplicant} for configuration of
802.11 networking, was updated to add experimental support for \ac{MAC} address
randomization in network scans.  Full implementation support was added in March
2015 \cite{suppl}.  Listing~\ref{lst:wpaSup} depicts the added support for
\ac{MAC} address randomization.  It is worth noting that the configuration file
provides two policies for using a non-globally unique address while in an
associated state. If the variable \texttt{mac\_addr} is set to \emph{1} the device
will use a randomized \ac{MAC} address for each unique network the device
connects to.  If \texttt{mac\_addr} is set to \emph{2} the device will randomize the
lower three bytes of the \ac{MAC} address prefixed with the original \ac{OUI}
where the local bit has been set to 1.

The \emph{wpa\_supplicant.conf} file also addresses the randomization policies
available for disassociated devices conducting active scanning.  In this case,
the variable \texttt{preassoc\_mac\_addr} can be set similarly to the previously
described address policies.

\begin{lstlisting}[caption=wpa\_supplicant.conf, label=lst:wpaSup]
# MAC address policy default
# 0 = use permanent MAC address
# 1 = use random MAC address for each ESS connection
# 2 = like 1, but maintain OUI (with local admin bit set)
#
# By default, permanent MAC address is used unless policy is changed by
# the per-network mac_addr parameter. Global mac_addr=1 can be used to
# change this default behavior.
#mac_addr=0

# Lifetime of random MAC address in seconds (default: 60)
#rand_addr_lifetime=60

# MAC address policy for pre-association operations (scanning, ANQP)
# 0 = use permanent MAC address
# 1 = use random MAC address
# 2 = like 1, but maintain OUI (with local admin bit set)
#preassoc_mac_addr=0
\end{lstlisting}


Android introduced \ac{MAC} address randomization for probe requests with
Android 6.0 (Marshmallow) and in an incremental patch to 5.0 (Lollipop).  With
the release of Marshmallow, the \emph{WifiStateMachine.java} and
\emph{WifiNative.java} files were modified to implement \ac{MAC} address
randomization for active scanning. When the \textit{SupplicantStartedState}
function is called upon enabling WiFi, a call to the newly added
\textit{setRandomMacOui} function sets the first three bytes of the \ac{MAC}
address to the default Google \ac{CID} (\texttt{DA:A1:19}).  If the
\texttt{config\_wifi\_random\_mac\_oui} variable has been redefined in the
\emph{config.xml} file, that prefix will be used in place of the default Google
\ac{CID}.  The XML configuration file allows an Android smartphone manufacturer
to override the default Google \ac{CID} with a prefix to be used as the
substitute for the \ac{OUI}.  Finally, the prefix is passed to another function,
\textit{setScanningMacOui} located in the \emph{WifiNative.java} file which
calls a corresponding function at a lower, native level. If the device chipset
is compatible to support randomization then the prefix will be used during
active scans.

We extracted the \emph{wpa\_supplicant.conf}, \emph{WifiStateMachine.java}, and
\emph{WifiNative.java} files from Android devices that do and do not perform
\ac{MAC} address randomization.  We found that the \emph{wpa\_supplicant} file
was never utilized to implement randomization, as attempts to modify the
randomization settings of the file had no affect on any device.  The Java files
also had the supporting functions for randomization included, regardless if the
device used them.  Interestingly, with logging enabled, the devices that did not
conduct randomization sent output to the logs indicating that the random
\ac{MAC} had been set, where devices seen randomizing did not.

\subsection{iOS}

In late 2014, Apple introduced \ac{MAC} address randomization with the
release of iOS 8.0.  Apple iOS randomization settings are not
device-model customizable, unlike Android, which allows each model to
modify settings such as the \ac{CID}.  As of the current iOS 10.x
version, Apple devices only use the locally assigned \ac{MAC} address
while in a disassociated state.  Since iOS is not open source, we
cannot determine the exact method or configuration options that Apple
uses on their devices to support randomization.  Instead, we are left
to determine device behavior from a ``black box'' perspective by
observing communication from different devices and iOS versions in
\S\ref{sec:analysis}.

\section{iOS Randomization Tests}
\label{randtests}
To determine if iOS is using random prefixes, or if there is just a pattern that
we have not been able to see, we used several standard statistical tests to
compare our observations with an ideal, random distribution.  First, we
calculated the number of \emph{collisions} we observed, where the same prefix
appeared more than once.  If they are truly random we would expect to see a
moderate number of collisions, which is easy to quantify.  We would also expect
to see a certain, far fewer, number of \emph{triple collisions} where one prefix
appears three times.  These numbers can be calculated as follows:

\vspace{-3mm}

\begin{align*}
  E[\text{\# of collisions}] &= \frac{\binom{n}{2}}{m} \\
  E[\text{\# of triple collisions}] &= \frac{\binom{n}{3}}{m^{2}} \\
  \text{where}~ n  &= \text{\# of addresses observed} \\ 
                m  &= \text{\# of possible prefixes ($2^{22}$)} \\ 
\end{align*}

\vspace{-3mm}
Comparing our empirical results with the statistical expectations, we get:
\vspace{-3mm}

\begin{align*}
  \text{For} &: \\
           \text{Collisions} &:  \text{expected = 266, observed = 262} \\
           \text{Triple collisions} &: \text{expected = 1, observed = 3}
\end{align*}

Additionally, we decomposed the bytes of subsequent \ac{MAC} addresses into a
bit stream and ran the tests specified in the FIPS 140-1 standard published by
NIST to test random number generators.  We obtained the following results:

\begin{itemize}
\item Monobit test: 9939
\item Poker test: 13.56
\item Runs test length 1: 2515 
\item Runs test length 2: 1342
\item Runs test length 3: 581
\item Runs test length 4: 281
\item Runs test length 5: 166
\item Longest run test: 12
\end{itemize}

All tests passed within the allowable ranges.  These tests indicate to us that
the \ac{MAC} addresses are distributed uniformly.  

\section{Google \ac{CID} Device Breakdown}
\label{da_breakdown}

\begin{table}[H]
  \centering
  \caption{\texttt{DA:A1:19} with \ac{WPS} Model Breakdown}
  \vspace{-2mm}
  \label{table:daModel}
  \scriptsize
  \begin{tabular}{ l | l | l }\hline 
  \textbf{Manufacturer} & \textbf{Model} & \textbf{Distinct Devices} \\ \hline \hline 
Huawei & Nexus 6P & 1660 \\ \hline
BlackBerry & STV100-3 & 133 \\ \hline
HTC & Nexus 9 & 107 \\ \hline
BlackBerry & STV100-1 & 71 \\ \hline
Sony & E5823 & 61 \\ \hline
Sony & E6653 & 59 \\ \hline
Sony & SO-01H & 29 \\ \hline
Sony & E6853 & 23 \\ \hline
Blackberry & STV100-4 & 20 \\ \hline
Huawei & NXT-L29 & 17 \\ \hline
Sony & SO-02H & 17 \\ \hline
Google & Pixel C & 12 \\ \hline
Sony & SO-03H & 11 \\ \hline
Sony & SOV32 & 11 \\ \hline
Huawei & NXT-AL10 & 11 \\ \hline
BlackBerry & STV100-2 & 10 \\ \hline
Sony & SO-03G & 9 \\ \hline
Sony & SOV31 & 8 \\ \hline
Sony & E6883 & 8 \\ \hline
Sony & E5803 & 8 \\ \hline
Sony & E6553 & 7 \\ \hline
Huawei & NXT-L09 & 6 \\ \hline
Sony & E6683 & 6 \\ \hline
Huawei & EVA-L09 & 5 \\ \hline
Sony & F5121 & 5 \\ \hline
Sony & E6533 & 4 \\ \hline
Huawei & EVA-AL00 & 3 \\ \hline
Huawei & KNT-AL20 & 2 \\ \hline
Huawei & EVA-AL10 & 2 \\ \hline
Sony & SGP712 & 2 \\ \hline
Sony & SGP771 & 2 \\ \hline
Sony & E6603 & 1 \\ \hline
Sony & E6633 & 1 \\ \hline
Sony & SO-05G & 1 \\ \hline
LGE & LG-H811 & 1 \\ \hline
Sony & E6833 & 1 \\ \hline
Huawei & VIE-AL10 & 1 \\ \hline
Huawei & EVA-DL00 & 1 \\ \hline
Sony & 402SO & 1 \\ \hline
Google & Pixel XL & 1 \\ \hline
Sony & 501SO & 1 \\ \hline
Huawei & EVA-L19 & 1 \\ \hline
Sony & F5321 & 1 \\ \hline
HTC & HTC 2PS650 & 1 \\ \hline 
  \end{tabular}
  
\end{table}
\clearpage
\section{\ac{RTS} Control Frame Attack - Device Diversity}
\label{rts_breakdown}

\begin{table}[h]
  \centering
  \caption{\ac{RTS} Control Frame Attack - Device Diversity}
  \vspace{-2mm}
  \label{table:activeAttack}
  \scriptsize{
  \begin{tabular}{ l | l | l }\hline 
    \textbf{Model} & \textbf{\ac{OS} Version} & \textbf{Success} \\ \hline \hline 
    iPhone 6s      & 10.1.1    &   $\surd$      \\ \hline
    iPhone 6s      & 9.3.5     &   $\surd$      \\ \hline
    iPhone 6s Plus & 9.3.5     &   $\surd$      \\ \hline
    iPhone 5s      & 10.1      &   $\surd$      \\ \hline
    iPhone 5s      & 9.3.5     &   $\surd$      \\ \hline
    iPhone 5       & 9.3.5     &   $\surd$      \\ \hline
    iPad Air       & 9.3.5     &   $\surd$      \\ \hline 
    Google Pixel XL& 7.1       &   $\surd$      \\ \hline 
    LGE Nexus 5X   & 7.0       &   $\surd$      \\ \hline
    LGE G5          & 6.0.1    &   $\surd$     \\ \hline
    LGE G4          & 6.0.1    &   $\surd$      \\ \hline
    Motorola Nexus 6 & 6.0.1   &   $\surd$      \\ \hline 
    Moto E2        &  5.1.1      &   $\surd$      \\ \hline 
    Moto Z Play     &  6.0.1   &   $\surd$      \\ \hline 
    OnePlus 3     &  6.0.1   &   $\surd$      \\ \hline 
    Xiaomi Mi Note Pro &  5.1.1   &   $\surd$      \\ \hline 

  \end{tabular}
}
\end{table}

\begin{acronym}
  \acro{AP}{access point}
  \acro{CDP}{Cisco Discovery Protocol}
  \acro{CID}{Company Identifier}
  \acro{CTS}{Clear-to-Send}
  \acro{DoS}{Denial-of-Service}
  \acro{DNS-SD}{Domain Name System-Based Service Discovery}
  \acro{GB}{gigabyte}
  \acro{GSM}{Global System for Mobile Communications}
  \acro{HPSW}{Hewlett-Packard SWitch Protocol}
  \acro{IE}{Information Element}
  \acro{IoT}{Internet of Things}
  \acro{IP}{Internet Protocol}
  \acro{IEEE}{Institute of Electrical and Electronics Engineers}
  \acro{IRB}{Institutional Review Board}
  \acro{LAN}{Local Area Network}
  \acro{MAC}{Media Access Control}
  \acro{MA-L}{MAC Address Block Large}
  \acro{MA-M}{MAC Address Block Medium}
  \acro{MA-S}{MAC Address Block Small}
  \acro{mDNS}{multicast Domain Name System}
  \acro{MNDP}{MicroTik Network Discovery Protocol}
  \acro{OS}{Operating System}
  \acro{OUI}{Organizationally Unique Identifier}
  \acro{P2P}{peer-to-peer}
  \acro{PC}{Personal Computer}
  \acro{PII}{Personally Identifiable Information}
  \acro{RF}{Radio Frequency}
  \acro{RTS}{Request-to-Send}
  \acro{SSID}{Service Set IDentifier}
  \acro{UUID-E}{Universally Unique IDentifier-Enrollee}
  \acro{VLAN}{Virtual Local Area Network}
  \acro{WPS}{Wi-Fi Protected Setup}

\end{acronym}

\end{document}